# Effect of Nanodiamond surfaces on Drug Delivery Systems

*Debsindhu Bhowmik[1,2,*], Gurpreet K. Dhindsa[1], Utsab R. Shrestha[1], Eugene Mamontov[3], and Xiang-qiang Chu[1,*]*

[1]*Dept. of Physics and Astronomy, Wayne State University USA;* [2]*Computational Sciences and Engineering Division, Oak Ridge National Laboratory,* [3]*Chemical and Engineering Materials Division, Oak Ridge National Laboratory*

*\*Email: bhowmikd@ornl.gov, chux@wayne.edu*

**Abstract:**

The prospect of RNA nanotechnology is increasing because of its numerous potential applications especially in medical science. The spherical Nanodiamonds (NDs) are becoming popular because of their lesser toxicity, desirable mechanical, optical properties, functionality and available surface areas. On other hand RNAs are stable, flexible and easy to bind to the NDs. In this work, we have studied the tRNA dynamics on ND surface by high-resolution quasi-elastic neutron scattering spectroscopy and all atom molecular dynamics simulation technique to understand how the tRNA motion is affected by the presence of ND. The flexibly of the tRNA is analyzed by the Mean square displacement analysis that shows tRNA have a sharp increase around 230K in its hydrated form. The intermediate scattering function (ISF) representing the tRNA dynamics follows the logarithmic decay as proposed by the Mode Coupling theory (MCT). But most importantly the tRNA dynamics is found to be faster in presence of ND within 220K to 310K compared to the freestanding ones. This is as we have shown, is because of the swollen RNA molecule due the introduction of hydrophilic ND surface.

**Introduction:**

Nanomaterials based on DNA are now well established[1], and have been in research for many drug delivery applications[2]. In parallel, the interest in RNA nanotechnology in recent years has been growing because of potential applications in medical science - from treating life-threatening diseases like cancer, genetic disorders to help cure untreatable viral illness[3]. Unlike DNA, the RNAs are more like proteins having flexibility[4–11], thermodynamically more stable and with some catalytic functions that can bind to functionalized nanoparticles such as Nanodiamond (ND). ND surface can be easily tailored with ionogenic groups (ether-C-O-C, peroxide –C-O-O-, carbonyl –C=O and hydroxyl type C-O-H bonding, etc.) and hydrocarbon fragments to adsorb large number of biologically active molecules[12–14]. Importantly, nearly spherical ND particles demonstrate no cellular toxicity and possess excellent mechanical and optical properties with easily available surfaces areas, functionality, and size (~5 nm diameter) close to that of intercellular structures and large biomolecules. It is believed that the spherical nanoparticles are good platform for the biological applications because of their simple geometry, consistency in shape and uniform surface chemistry. They are safer compared to tubular shaped nanomaterials because of their toxicological properties. RNAs can be folded into various complex structures to be used later as a platform for nanomedicine applications. Consequently the RNA-ND complex is well suited for biomedical applications - such as drug-delivery, tissue engineering, tribology and bio-imaging. .

Many small and biologically active molecules have been linked or adsorbed on NDs. An important question is, however, to what extent the ND linked/adsorbed molecules will retain their biological activity? This question is especially relevant for proteins and nucleic acids for which the activity strongly depends on their molecular conformations, which in turn are very sensitive to the environment and have been demonstrated to change upon adsorption[15,16]. Therefore, understanding the details of biomacromolecules interactions with the surface of the nanoparticles is of great importance for design of nanocarriers for proteins, DNA, and RNA.

Towards this aim, in the present work we intend to explore the micro dynamics of hydrated and dry tRNA molecules on ND surface relative to the freestanding ones. The quasi-elastic neutron scattering (QENS) technique which is an appropriate tool to track atomic level dynamics on such a length scale not only for biomacromolecules but also for polyelectrolytes, polymers and soft nanocomposites[4–11,17–27] - from Å to nm with pico to nanosecond time window has been used here; while the activation or flexibility - a universal property exhibited by the proteins[28] and biopolymers[29] - is measured by elastic neutron Scattering. The use of neutrons is beneficial because of its unique property to specially track the hydrogen atoms, as the incoherent scattering cross-section of hydrogen atom is very large compared to others. Thus by replacing water to heavy water, we make sure that only tRNAs contain hydrogen and then we unambiguously follow the tRNA dynamics. Additionally we make use of Molecular Dynamics (MD) simulation, a complementary and effective method to extract information that in many cases is difficult or impossible to attain experimentally[21,23,25]. Somewhat unexpectedly, within the range of temperatures investigated, our results distinctly show a faster tRNA dynamics on ND surface compared to dry or hydrated RNA molecules. This is in a striking contrast to, for instance, adsorption of myoglobin on silica surface, where the protein dynamics slowed down compared to dissolved state – a finding explained by the authors due to simultaneous decrease in flexibility and the prevailing modification of the mobility of residues and dynamics of the protein upon its interaction with surface[30]. Whereas in our work, the faster tRNA motion in presence of ND surface is because of cage formation by faster water molecules due to the introduction of ND into the system.

**Results and analysis:**

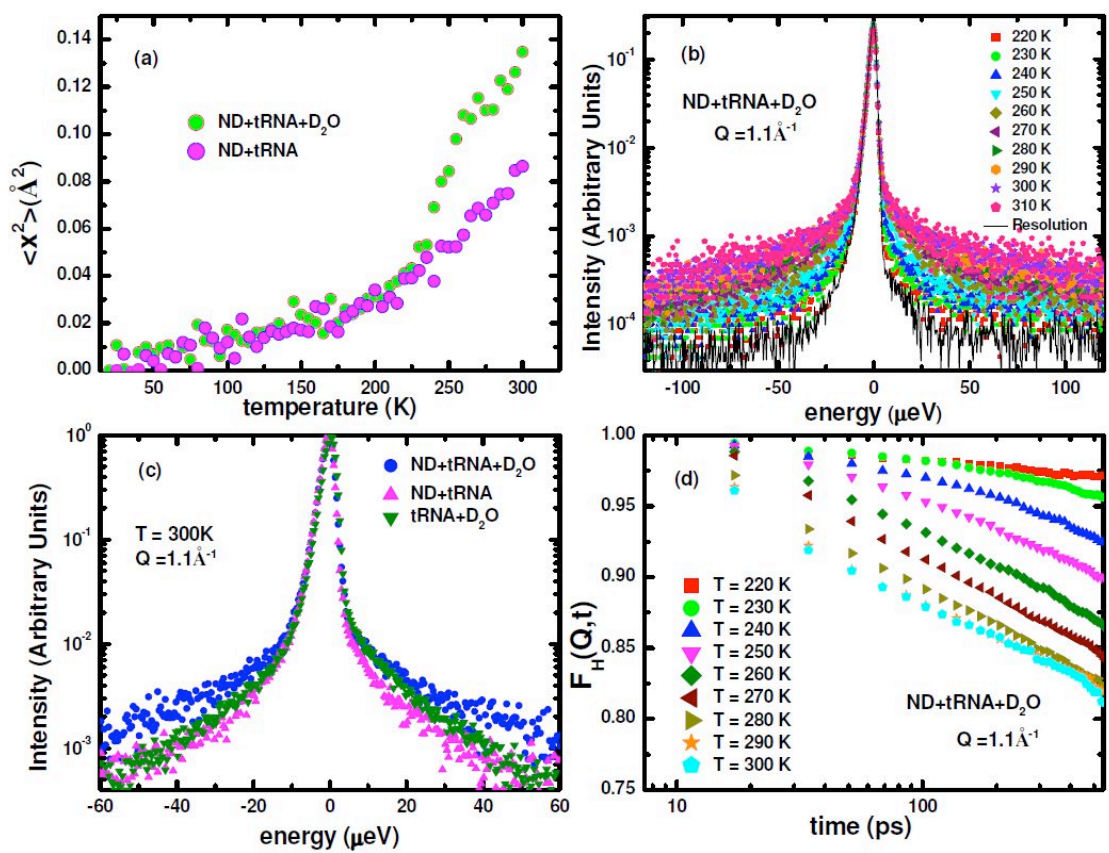

*Figure 1: (a) MSD, $<x^2(T)>$ of $ND+tRNA+D_2O$ and $ND+tRNA$ as a function of temperature; (b) $S(Q,\omega)$ of $ND+tRNA+D_2O$ at different temperatures from 220K-310K at $Q=1.1Å^{-1}$; (c) Comparison of $ND+tRNA+D_2O$, $ND+tRNA$ and $tRNA+D_2O$ in energy domain at 300K ($Q=1.1Å^{-1}$); (d) Normalized intermediate scattering function (Fourier transformed $S(Q,\omega)$ data in panel Figure 1(b)) of $ND+tRNA+D_2O$ system.*

***Mean Square displacement:*** Similar to proteins[8], the 'softness' or 'flexibility' of an RNA molecule is usually measured by incoherent elastic scattering and calculated through Mean Square Displacement (MSD) analysis using Debye Waller Factor, $S(Q,\omega=0,T)/S(Q,\omega=0,T=4K) = \exp[-Q^2<x^2(T)>]$. MSD - $<x^2(T)>$ of hydrated and dry tRNA with ND, plotted in Fig. 1(a) as a function of temperature. Three onsets are seen in the case of hydrated RNA system (with ND). The first onset occurs around 100K both in the hydrated and dry sample. It originates from the rotational motion of hydration independent methyl groups[31] and can be observed in a number of proteins[32,33]. A second one, at ~230K, is noticed in hydrated system but not in the dry one. This inflection in the slope is sometimes called as 'dynamic transition' – that is responsible for the fluctuation of backbone and the side chain activation which are generally hydrophilic in nature. This is hydration dependent and thus not seen in dry sample[34,35]. On the other hand, proteins or nucleic acids dynamics have been found to follow glassy dynamics[36,37] and this glassy behavior can be observed both in dry and in hydrated samples around 170K.

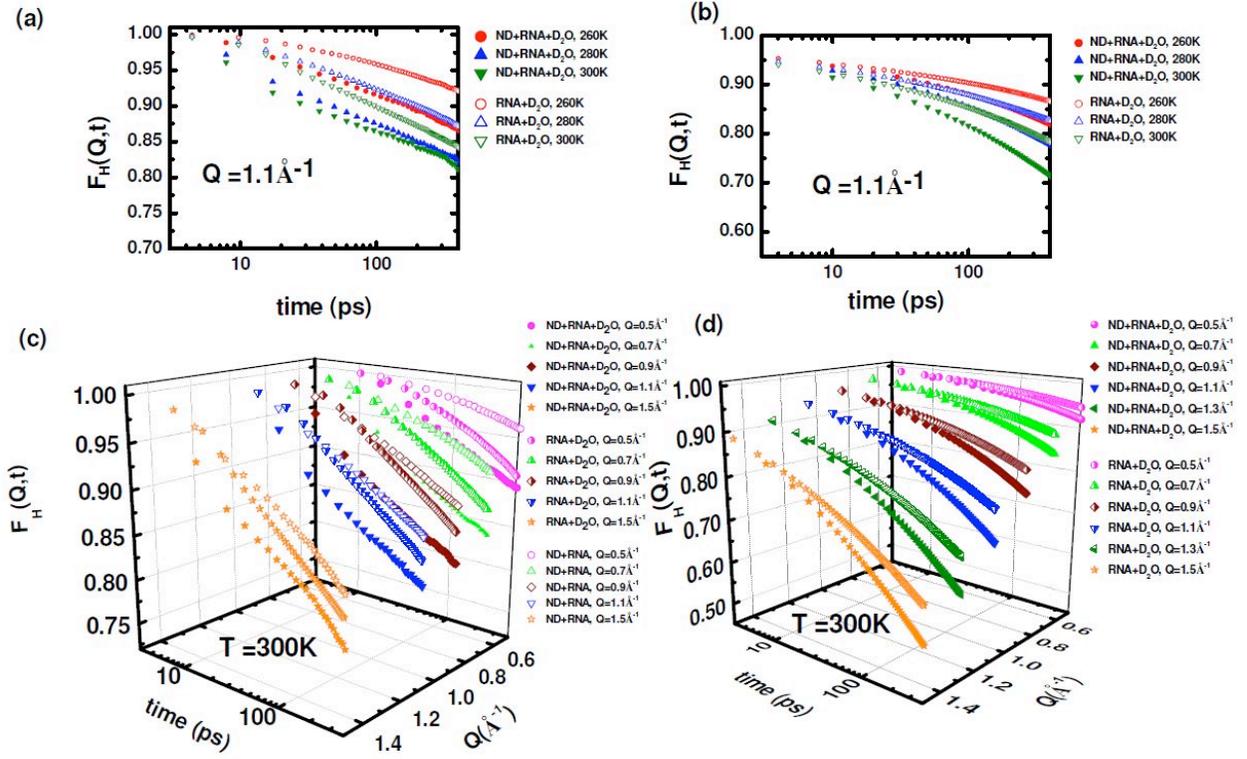

*Figure 2: (a) and (b) Comparison of experimental and MD Simulated ISF of ND+tRNA+D$_2$O and tRNA+D$_2$O at three different temperatures (Q= 1.1Å$^{-1}$), respectively. (c) and (d) Comparison of experimental and MD Simulated ISF of ND+tRNA+D$_2$O, ND+tRNA and tRNA+D$_2$O at five different length scales ranging from 0.5Å$^{-1}$ to 1.5Å$^{-1}$ and at 300K.*

***QENS data in energy domain:*** The QENS experiments provide the dynamic structure factor $S(Q,\omega)$. In Figure 1(b) we show the normalized $S(Q,\omega)$ of D$_2$O hydrated tRNA on ND surface at Q= 1.1Å$^{-1}$ between 220K to 310K revealing how the motion intensifies with the temperature. As mentioned before, it should be noted that the motion captured in the QENS spectra is basically the average motion of hydrogen atoms present in the system. The uniqueness of neutrons compared, for example, to X-Rays is that the neutrons are most sensitive to protons because of their unusually high incoherent neutron scattering length compared to other atoms, including deuterium. The broadening of QENS spectra about the center of the bell-shaped $S(Q,\omega)$ curves demonstrates the average hydrogen dynamics within the spectrometer window. Faster motion results in broader peaks. In Figure 1(c) all three different systems (D$_2$O hydrated tRNA, dry tRNA on ND and D$_2$O hydrated tRNA on ND) at the ambient temperature, are plotted together. The direct comparison of the broadenings among the $S(Q,\omega)$ spectra of these three different systems clearly points that the hydrogens of tRNA dissolved in $D_2O$ have faster dynamics in presence of the ND in the solution compared to the other sample.

***ISF in time domain:*** In Figure 1(d), the intermediate scattering functions, $F_H(Q,t)$ are shown using the Fourier transformation of the Figure 1(b) results. A faster decay in $F_H(Q,t)$ signifies a faster dynamics. A direct comparison of the $F_H(Q,t)$ decay in hydrated RNA with and without the ND at different temperature is shown in Figure 2(a), where a faster RNA dynamics is visible in presence of ND. Similarly from the Figure 2(c), it is clear that the RNA is faster in presence of ND than without ND or dry RNA over all the length scale (0.5-1.5Å$^{-1}$ ~ Å to nm range).

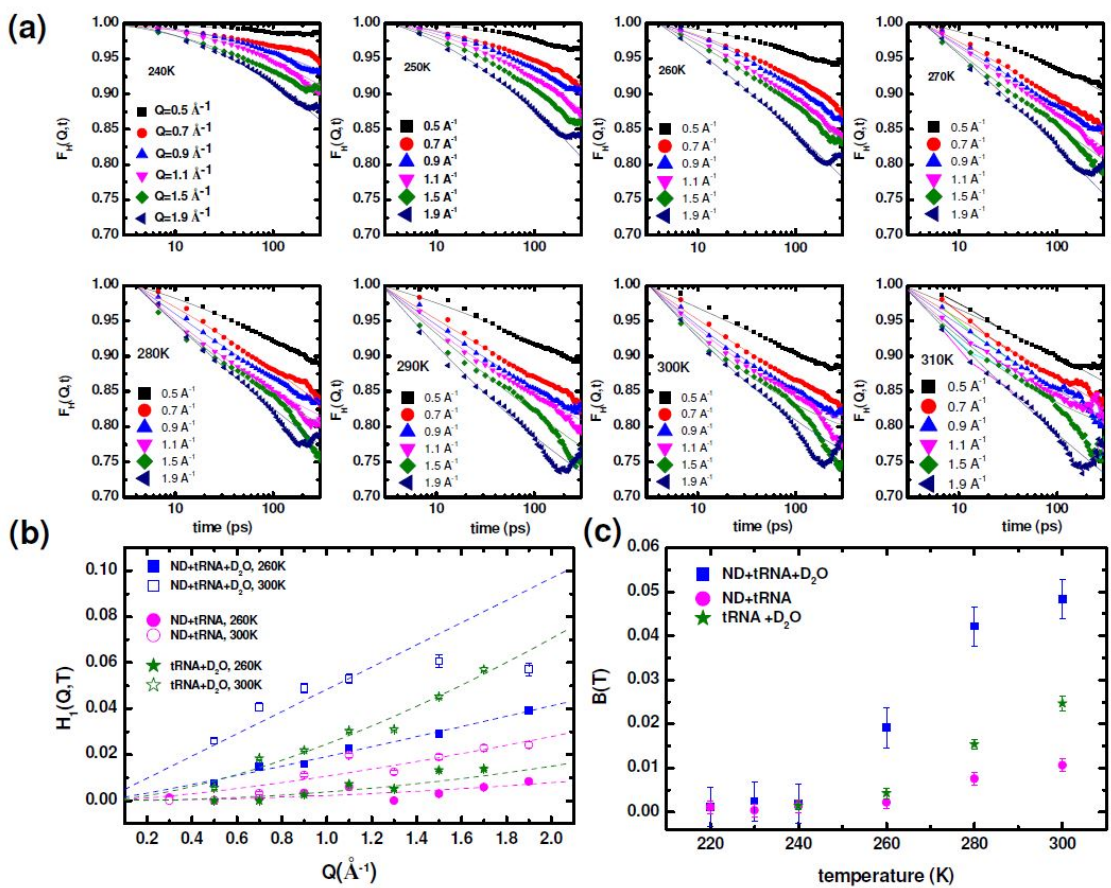

*Figure 3: (a) Experimental Intermediate Scattering function from $D_2O$ hydrated tRNA on ND surface, are fitted with logarithmic model (eq. 2)[28,36,38,39] at different Q values (covering ~3Å to ~1.3nm) and from 240K to 310K; (b) and (c) $H_1(Q,T)$ and $B(T)$ are fitted parameters of logarithmic model (Eq. 2). $H_1(Q,T)$ plotted as a function of Q at three different temperatures while $B(T)$ plotted as a function of temperature.*

All-atom MD simulations[24,25] are performed to help interpret the faster dynamics of the D$_2$O hydrated tRNA on ND surface as revealed by the QENS experiments. The simulated normalized ISFs of RNA hydrogens (D$_2$O hydrated) calculated at different temperatures are presented in Figure 2(b). Similar to the QENS experiments, the simulated ISFs also confirm a faster dynamics of the D$_2$O hydrated RNA in the presence of ND. Further in Figure 1(d), we have shown the simulated $F_H(Q,t)$ at different length scale (0.5Å$^{-1}$-1.5Å$^{-1}$) at ambient temperature similar to Figure 1(c). Our results show both experimentally and confirmed by simulation that tRNA becomes faster in its hydrated form when ND is present.

***Time domain QENS data analysis:*** As we have previously showed the D$_2$O hydrated tRNA follows the logarithmic relaxation dynamics as a function of time[29], we applied the same Mode Coupling Theory (MCT) to the D$_2$O hydrated tRNA on ND. This says $F_H(Q,t)$ can be fitted by an asymptotic expression derived from Mode Coupling Theory (MCT)[37–39]:

$$F_H(Q,t) \sim [f(Q,T) - H_1(Q,T)\ln(t/\tau_\beta(T)) + H_2(Q,T)\ln^2(t/\tau_\beta(T))]\exp(t/\tau_\alpha(Q,T)) \quad (1)$$

where $\tau_\beta(T)$ and $\tau_\alpha(Q,T)$ are the characteristic β- and α- relaxation times, $f(Q,T)$ is a temperature-dependent pre-factor, which is proportional to the Debye-Waller factor for small Qs, i.e. $f(Q,T) = \exp[-A(T)Q^2]$. The Q- dependent parameters, $H_1(Q,T)$ and $H_2(Q,T)$, can be written as $H_1(Q,T) = h_1(Q)B_1(T)$ and $H_2(Q,T) = h_1(Q)B_2(Q,T)$, representing the first and second order logarithmic decay parameter[37]. In our experiment, the time range (up to 1 ns) is much shorter than α-relaxation time range (μs to ms), hence the value of the last exponential factor can be approximated to unity. In that case equation (5) can be simplified as

$$F_H(Q,t) \sim [f(Q,T) - H_1(Q,T)\ln(t/\tau_\beta(T)) + H_2(Q,T)\ln^2(t/\tau_\beta(T))] \quad (2)$$

The ISF of the average of hydrogen in D$_2$O hydrated tRNA on ND surface can be well explained according to equation 2 between 240K to 310K and within 0.3Å$^{-1}$ to 1.9Å$^{-1}$ (~3 Å to ~2nm) (Figure 3(a)). The four parameters in this model, $A(T)$, $\tau_\beta(T)$, $H_1(Q,T)$ and $H_2(Q,T)$ are obtained by fitting all six different Q curves simultaneously at each temperature. $A(T)$ is fixed at zero because the pre-factor $f(Q,T)$ goes to 1 at all the Q-values at a specific short time $\tau\beta(T)$ ~10 ps which is much shorter than our measured time range. Figure 3(b) represents $H_1(Q,T)$ of dry and hydrated tRNA on ND, that follows a power law at small Q according to $H_1(Q,T) = B_1(T)Q^\beta$, with β between 1 and 2, and $B_1(T)$, a temperature dependent parameter, is plotted in Figure 3(c). From the $H_1(Q,T)$ and $B_1(T)$ values among the three different systems (Figure 3(b) and 3(c)); a faster decay in the tRNA dynamics (D$_2$O hydrated) can be quantified in presence of ND over the others at all the temperatures and over the length scale studied (from 0.3Å$^{-1}$ to 1.9Å$^{-1}$ i.e ~3Å-2nm).

## Discussion:

From our MSD results we have not observed any ice formation, generally exhibited by sharp drop in the MSD curves, within the temperature range studied. The three onsets observed in the $D_2O$ hydrated tRNA on surfaces denotes three different situations. The first one at ~100K is an inflection point denoting the anharmonic[30,35] behavior due to the local methyl group rotation[31] and commonly seen in many protein systems[32,33]. This is hydration independent and thus can be noticed also in the dry sample. Until the temperature reaches ~230K, hydration does not play any major role[31,32,40,41]. The transition at ~230K is a manifestation of the backbone fluctuations[42,43] and the activation of the hydrophilic side chains due to the hydrogen bond formation and its relaxation with the surrounding water molecules[44,45]. This is in sometimes called as the 'dynamic transition' and might be responsible for triggering the biological functions. This also assures no denaturation of the $D_2O$ hydrated tRNA because of the absorption on ND surface. As it is hydration dependent[30] obviously then it cannot be seen in the dry sample because of the lack of hydration water[34,35]. The other inflection point which is sometimes referred to as the glassy behavior is seen ~170K in both the dry and the $D_2O$ hydrated tRNA on ND surface. The glass transition temperature is typically lower than dynamic transition temperature that has been attributed to the caged dynamics caused by strong intermolecular interactions and responsible for the existence of dynamic transition in proteins[46].

Coming to the study of micro-dynamics the obvious question is what causes this faster dynamics in hydrated tRNA in presence of ND? - The finding that we confirmed both by our QENS experiment and MD simulation. In brief, both water and tRNA stay very close to the ND surface that is hydrophilic in nature and water molecules forms distinct shell around ND and tRNA molecules. The water shell around ND surface prevents outer water to come closer and hydrophilic ND surface significantly modifies tRNA dynamics. Besides there is almost an order of magnitude more water molecules around ND than that around tRNA. Because tRNA is itself hydrophilic with high presence near ND surface this augments tRNA hydration level that in turn causes tRNA swelling. Consequently this leads to faster dynamics of tRNA in presence of ND surface as explained in detail with MD simulation results[25].

## Conclusion:

In summary we have studied the microscopic dynamics of tRNA on ND surfaces and compared it with the dry tRNA dynamics on ND surface and with hydrated tRNA without ND both in energy and in time domain. While analyzing the temperature dependent MSD data, we have not noticed three onsets on the hydrated tRNA samples on ND surface while for dry sample on ND, only two are observed. The first two inflection points ~100K and ~170K are due to the local methyl group rotation independent of hydration effect and because of caged dynamics at the glass transition temperature. Both of these are visible in hydrated and dry tRNA on ND surface. The third one at around 230K is hydration dependent and thus be observed only in the hydrated tRNA on ND and due to the backbone fluctuation and activation of side chains by forming hydrogen bond with surrounding water molecules.

From both the energy and time domain analysis it can be said that the hydrated tRNA dynamics becomes faster on ND surface compared to dry tRNA and hydrated tRNA without ND. The larger broadening in the $S(Q,\omega)$ spectra and the faster decay in the $F_H(Q,t)$ in the $D_2O$ hydrated tRNA on ND surface over ~3Å-2nm (i.e. $0.3Å^{-1}$ to $1.9Å^{-1}$) confirms that tRNA dynamics becomes faster in presence of ND. Further the logarithm decay as predicted by the Mode Coupling Theory (MCT) which has been also seen in other proteins and biological macromolecules including tRNA, is found to be applicable here. But the answer to the rather surprising finding about why the ND presence makes the hydrated tRNA dynamics faster is that the water shell around ND increases tRNA hydration level around ND leading to its swelling and increases dynamics. In short it has been shown that the large biopolymers or biomolecules like tRNA can well retain its folded state and activate biological function even in presence of hydrophilic ND surface but simultaneously attain a faster motion compared to the dry or freestanding tRNA.

**Materials and Methods:**

NDs (radius ~2.5nm) were prepared with detonation technique[13] and were placed inside the vacuum to remain in dry condition. tRNA was bought from Sigma Aldrich and was used without further purification and lyophilization.. The tRNA was adsorbed on 0.33g of ND surface and two samples were prepared. One of them was hydrated with $D_2O$ and the other one was dry - (a) ND+tRNA+$D_2O$, (b) ND+tRNA, respectively. The hydrated sample was prepared by adsorbing the 0.03 gm of tRNA on 0.33g on dry ND surface, and then hydrated with 0.06g of $D_2O$. In the dry sample, 0.1gm of tRNA was adsorbed on the ND surface. Because a considerable portion of water goes to ND surface, it can not be properly calculated what is the exact hydration level of tRNA in the system. Nevertheless from MD simulation we have checked the number of water molecules around the RNA remain same when ND is present in the system compared to without ND (Figure 5(c) and Figure 5(d)). Also for tRNA it is known to notice crystallization of water if the hydration level goes above 0.65h (g of $D_2O$/g of tRNA) and we have not noticed any indication of crystallization[47]. For $D_2O$ hydrated tRNA sample without ND, we used data from our old experiment where tRNA hydration level was 0.5h (g of $D_2O$/g of tRNA)[29].

**References:**

(1)    Aldaye, F. a; Palmer, A. L.; Sleiman, H. F. Assembling Materials with DNA as the Guide. *Science* **2008**, *321*, 1795–1799.

(2)    Seeman, N. C. Nanomaterials Based on DNA. *Annu. Rev. Biochem.* **2010**, *79*, 65–87.

(3)    Guo, P. The Emerging Field of RNA Nanotechnology. *Nat. Nanotechnol.* **2010**, *5*, 833–842.

(4) Perera, S. M. D. C.; Shrestha, U.; Bhowmik, D.; Chawla, U.; Struts, A. V.; Chu, X.; Brown, M. F. Neutron Scattering Reveals Protein Fluctuations in GPCR Activation. *Biophys. J.* **2016**, *110*, 228a–229a.

(5) Bhowmik, D.; Shrestha, U.; Perera, S. M. d. c.; Chawla, U.; Mamontov, E.; Brown, M. F.; Chu, X.-Q. Rhodopsin Photoactivation Dynamics Revealed by Quasi-Elastic Neutron Scattering. *APS March Meet. 2015* **2015**.

(6) Bhowmik, D.; Kaur Dhindsa, G.; Rusek, A. J.; Delinder, K. Van; Shrestha, U.; Ng, J. D.; Sharp, M.; Stingaciu, L. R.; Chu, X. Probing the Domain Motions of an Oligomeric Protein from Deep-Sea Hyperthermophile by Neutron Spin Echo. *Biophys. J.* **2015**, *108*, 59a.

(7) Shrestha, U. R.; Bhowmik, D.; Copley, J. R. D.; Tyagi, M.; Leao, J. B.; Chu, X.-Q. Effects of Pressure on the Dynamics of a Hyperthermophilic Protein Revealed by Quasielastic Neutron Scattering. *APS March Meet. 2016* **2016**.

(8) Shrestha, U. R.; Bhowmik, D.; Copley, J. R. D.; Tyagi, M.; Leão, J. B.; Chu, X.-Q.; Klein, M. L. Effects of Pressure on the Dynamics of an Oligomeric Protein from Deep-Sea Hyperthermophile. *Proc. Natl. Acad. Sci. U. S. A.* **2015**, *112*, 13886–13891.

(9) Shrestha, U.; Bhowmik, D.; Perera, S. M. D. C.; Chawla, U.; Struts, A. V.; Graziano, V.; Pingali, S. V.; Heller, W. T.; Qian, S.; Brown, M. F.; et al. Small-Angle Neutron and X-ray Scattering Reveal Conformational Changes in Rhodopsin. *APS March Meet. 2015* **2015**, 1–2.

(10) Bhowmik, D.; Shrestha, U.; Perera, S. M. d. c.; Chawla, U.; Mamontov, E.; Brown, M. F.; Chu, X.-Q. Rhodopsin Photoactivation Dynamics Revealed by Quasi-Elastic Neutron Scattering. *Biophys. J.* **2015**, *108*, 61a.

(11) Bhowmik, D.; Shrestha, U. R.; Dhindsa, G. K.; Sharp, M.; Stingaciu, L. R.; Chu, X. Slow Domain Motions of an Oligomeric Protein from Deep-Sea Hyperthermophile Probed by Neutron Spin Echo. *APS March Meet. 2016* **2016**.

(12) Chung, P. H.; Perevedentseva, E.; Tu, J. S.; Chang, C. C.; Cheng, C. L. Spectroscopic Study of Bio-Functionalized Nanodiamonds. *Diam. Relat. Mater.* **2006**, *15*, 622–625.

(13) Kharlamova, M. V.; Mochalin, V. N.; Lukatskaya, M. R.; Niu, J.; Presser, V.; Mikhalovsky, S.; Gogotsi, Y. Adsorption of Proteins in Channels of Carbon Nanotubes: Effect of Surface Chemistry. *Mater. Express* **2013**, *3*, 1–10.

(14) Liu, K.-K.; Zheng, W.-W.; Wang, C.-C.; Chiu, Y.-C.; Cheng, C.-L.; Lo, Y.-S.; Chen, C.; Chao, J.-I. Covalent Linkage of Nanodiamond-Paclitaxel for Drug Delivery and Cancer Therapy. *Nanotechnology* **2010**, *21*, 315106.

(15) Wang, J.; Hu, Z.; Xu, J.; Zhao, Y. Therapeutic Applications of Low-Toxicity Spherical Nanocarbon Materials. *NPG Asia Mater.* **2014**, *6*, e84.

(16) Chu, X.; Gajapathy, M.; Weiss, K. L.; Mamontov, E.; Ng, J. D.; Coates, L. Dynamic Behavior of Oligomeric Inorganic Pyrophosphatase Explored by Quasielastic Neutron Scattering. *J. Phys. Chem. B* **2012**, *116*, 9917–9921.

(17) Arbe, A.; Pomposo, J. A.; Asenjo-Sanz, I.; Bhowmik, D.; Ivanova, O.; Kohlbrecher, J.; Colmenero, J. Single Chain Dynamic Structure Factor of Linear Polymers in an All-Polymer Nano-Composite. *Macromolecules* **2016**, *49*, 2354–2364.

(18) Bhowmik, D.; Pomposo, J. A.; Juranyi, F.; Sakai, V. G.; Zamponi, M.; Su, Y.; Arbe, A.; Colmenero, J. Microscopic Dynamics in Nanocomposites of Poly(ethylene Oxide) and Poly(methyl Methacrylate) Soft Nanoparticles: A Quasi-Elastic Neutron Scattering Study. *Macromolecules* **2014**, *47*, 304–315.

(19) Bhowmik, D.; Pomposo, J. A.; Juranyi, F.; Sakai, V. G.; Zamponi, M.; Arbe, A.; Colmenero, J. Investigation of a Nanocomposite of 75 Wt % Poly ( Methyl Methacrylate ) Nanoparticles with 25 Wt % Poly ( Ethylene Oxide ) Linear Chains : A Quasielatic Neutron Scattering , Calorimetric , and WAXS Study. *Macromolecules* **2014**, *47*, 3005–3016.

(20) Bhowmik, D.; Pomposo, J. A.; Juranyi, F.; Sakai, V. G.; Zamponi, M.; Su, Y.; Arbe, A.; Colmenero, J. *Quasielastic Neutron Scattering Insight into the Molecular Dynamics of All-Polymer Nano-Composites*; Munich, 2015.

(21) Bhowmik, D.; Malikova, N.; Teixeira, J.; Mériguet, G.; Bernard, O.; Turq, P.; Häussler, W. Study of Tetrabutylammonium Bromide in Aqueous Solution by Neutron Scattering. *Eur. Phys. J. Spec. Top.* **2012**, *213*, 303–312.

(22) Bhowmik, D. Study of Microscopic Dynamics of Complex FLuids Containing Charged Hydrophobic Species by Neutron Scattering Coupled with Molecular Dynamics Simulation, UNIVERSITE PARIS VI PIERRE ET MARIE CURIE (UNIVERSITE PARIS VI), 2011.

(23) Bhowmik, D.; Malikova, N.; Mériguet, G.; Bernard, O.; Teixeira, J.; Turq, P. Aqueous Solutions of Tetraalkylammonium Halides: Ion Hydration, Dynamics and Ion-Ion Interactions in Light of Steric Effects. *Phys. Chem. Chem. Phys.* **2014**, *16*, 13447–13457.

(24) Dhindsa, G. K.; Bhowmik, D.; Ganesh, P.; Goswami, M.; Mochalin, V. N.; O'Neill, H.; Gogotsi, Y.; Mamontov, E.; Chu, X. Effect of Nanodiamond Surfaces on tRNA Dynamics Studied by Neutron Scattering an. *APS March Meet. 2015* **2015**, 1–2.

(25) Dhindsa, G. K.; Bhowmik, D.; Goswami, M.; O'Neill, H.; Mamontov, E.; Sumpter, B. G.; Hong, L.; Panchapakesan, G.; Chu, X. Enhanced Dynamics of Hydrated tRNA on Nanodiamond Surfaces : A Combined Neutron Scattering and MD Simulation Study. *J. Phys. Chem. B* **2016**, *(in press)*, doi:10.1021/acs.jpcb.6b07511.

(26) Malikova, N.; Čebašek, S.; Glenisson, V.; Bhowmik, D.; Carrot, G.; Vlachy, V. Aqueous Solutions of Ionenes: Interactions and Counterion Specific Effects as Seen by Neutron Scattering. *Phys. Chem. Chem. Phys.* **2012**, *14*, 12898–12904.

(27) Shrestha, U.; Bhowmik, D.; Perera, S. M. D. C.; Chawla, U.; Struts, A. V.; Graziano, V.; Qian, S.; Heller, W. T.; Brown, M. F.; Chu, X.-Q. Small Angle Neutron and X-Ray Scattering Reveal Conformational Differences in Detergents Affecting Rhodopsin Activation. *Biophys. J.* **2015**, *108*, 39a.

(28) Chu, X.; Mamontov, E.; O'Neill, H.; Zhang, Q. Temperature Dependence of Logarithmic-like Relaxational Dynamics of Hydrated tRNA. *J. Phys. Chem. Lett.* **2013**, *4*, 936–942.

(29) Devineau, S.; Zanotti, J. M.; Loupiac, C.; Zargarian, L.; Neiers, F.; Pin, S.; Renault, J. P. Myoglobin on Silica: A Case Study of the Impact of Adsorption on Protein Structure and Dynamics. *Langmuir* **2013**, *29*, 13465–13472.

(30) Roh, J. H.; Curtis, J. E.; Azzam, S.; Novikov, V. N.; Peral, I.; Chowdhuri, Z.; Gregory, R.

B.; Sokolov, a P. Influence of Hydration on the Dynamics of Lysozyme. *Biophys. J.* **2006**, *91*, 2573–2588.

(31) Roh, J. H.; Novikov, V. N.; Gregory, R. B.; Curtis, J. E.; Chowdhuri, Z.; Sokolov, a. P. Onsets of Anharmonicity in Protein Dynamics. *Phys. Rev. Lett.* **2005**, *95*, 38101.

(32) Telling, M. T. F.; Howells, S.; Combet, J.; Clifton, L. a.; García Sakai, V. Mean Squared Displacement Analysis of an-Harmonic Behaviour in Lyophilised Proteins. *Chem. Phys.* **2013**, *424*, 32–36.

(33) Caliskan, G.; Briber, R. M.; Thirumalai, D.; Garcia-Sakai, V.; Woodson, S. a.; Sokolov, A. P. Dynamic Transition in tRNA Is Solvent Induced. *J. Am. Chem. Soc.* **2006**, *128*, 32–33.

(34) Dhindsa, G. K.; Tyagi, M.; Chu, X.-Q. Temperature-Dependent Dynamics of Dry and Hydrated β-Casein Studied by Quasielastic Neutron Scattering. *J. Phys. Chem. B* **2014**, *118*, 10821–10829.

(35) Doster, W.; Cusack, S.; Petry, W. Dynamical Transition of Myoglobin Revealed by Inelastic Neutron Scattering. *Nature* **1989**, *337*, 754–756.

(36) Lagi, M.; Baglioni, P.; Chen, S.-H. Logarithmic Decay in Single-Particle Relaxation of Hydrated Lysozyme Powder. *Phys. Rev. Lett.* **2009**, *103*, 108102.

(37) Ngai, K. L.; Capaccioli, S.; Paciaroni, a. Change of Caged Dynamics at Tgin Hydrated Proteins: Trend of Mean Squared Displacements after Correcting for the Methyl-Group Rotation Contribution. *J. Chem. Phys.* **2013**, *138*.

(38) Sciortino, F.; Tartaglia, P.; Zaccarelli, E. Evidence of a Higher-Order Singularity in Dense Short-Ranged Attractive Colloids. *Phys. Rev. Lett.* **2003**, *91*, 268301.

(39) Chu, X.; Lagi, M.; Mamontov, E.; Fratini, E.; Baglioni, P.; Chen, S.-H. Experimental Evidence of Logarithmic Relaxation in Single-Particle Dynamics of Hydrated Protein Molecules. *Soft Matter* **2010**, *6*, 2623.

(40) Wood, K.; Tobias, D. J.; Kessler, B.; Gabel, F.; Oesterhelt, D.; Mulder, F. a a; Zaccai, G.; Weik, M. The Low-Temperature Inflection Observed in Neutron Scattering Measurements of Proteins Is due to Methyl Rotation: Direct Evidence Using Isotope Labeling and Molecular Dynamics Simulations. *J. Am. Chem. Soc.* **2010**, *132*, 4990–4991.

(41) Doster, W.; Settles, M. Protein-Water Displacement Distributions. *Biochim. Biophys. Acta - Proteins Proteomics* **2005**, *1749*, 173–186.

(42) Schiró, G.; Caronna, C.; Natali, F.; Cupane, A. Molecular Origin and Hydration Dependence of Protein Anharmonicity: An Elastic Neutron Scattering Study. *Phys. Chem. Chem. Phys.* **2010**, *12*, 10215–10220.

(43) Schiró, G.; Caronna, C.; Natali, F.; Cupane, A. Direct Evidence of the Amino Acid Side Chain and Backbone Contributions to Protein Anharmonicity. *J. Am. Chem. Soc.* **2010**, *132*, 1371–1376.

(44) Miao, Y.; Yi, Z.; Glass, D. C.; Hong, L.; Tyagi, M.; Baudry, J. Temperature-Dependent Dynamical Transitions of Di Ff Erent Classes of Amino Acid Residue in a Globular Protein. *J. Am. Chem. Soc.* **2012**, *134*, 19576–19579.

(45) Tournier, A. L.; Xu, J.; Smith, J. C. Translational Hydration Water Dynamics Drives the

Protein Glass Transition. *Biophys. J.* **2003**, *85*, 1871–1875.

(46) Frontzek, A. V.; Strokov, S. V.; Embs, J. P.; Lushnikov, S. G. Does a Dry Protein Undergo a Glass Transition? *J. Phys. Chem. B* **2014**, *118*, 2796–2802.

(47) Roh, J. H.; Briber, R. M.; Damjanovic, a; Thirumalai, D.; Woodson, S. a; Sokolov, a P. Dynamics of tRNA at Different Levels of Hydration. *Biophys. J.* **2009**, *96*, 2755–2762.